\documentclass[%twocolumn,
reprint,
 amsmath,amssymb,
aps,
prl
]{revtex4-1}

\usepackage[dvipdfmx]{graphicx}% Include figure files
\usepackage{dcolumn}% Align table columns on decimal point
\usepackage{bm}% bold math
\usepackage{hyperref}% add hypertext capabilities

\usepackage{upgreek}

\usepackage{mathptmx}

\usepackage{cancel}

\begin{document}

\preprint{AIP/123-QED}

\title{Intrinsic and extrinsic tunability of Rashba spin-orbit coupled emergent inductors}
% Force line breaks with \\
\author{Jun'ichi Ieda$^{1,2}$ and Yuta Yamane$^{2,3}$}
\affiliation{$^1$Advanced Science Research Center, Japan Atomic Energy Agency, Tokai 319-1195, Japan}
\affiliation{$^2$Research Institute of Electrical Communication, Tohoku University, Sendai 980-8577, Japan}
\affiliation{$^3$Frontier Research Institute for Interdisciplinary Sciences, Tohoku University, Sendai 980-8578, Japan}

\date{\today}% It is always \today, today,

\begin{abstract}
The emergent induction of spiral magnets that was proposed [Jpn.~J.~Appl.~Phys.~{\bf 58}, 120909 (2019)] and recently demonstrated [Nature {\bf 586}, 232 (2020)] is shown to be further extended by a comprehensive treatment of the Rashba spin-orbit coupling and the electron spin relaxation that affect the underlying processes of spin-transfer torque and spinmotive force. Within adiabatic approximation, we show that the output voltages are widely altered intrinsically via the Rashba effect whereas extrinsically via the nonadiabatic correction due to the spin relaxation and sample disorder. The findings respectively clarify the origins for the amplitude modulation and sign change of the emergent inductance with tunability by electrical gating and careful sample preparation.
\end{abstract}

\maketitle

%\begin{quotation}
%The ``lead paragraph'' is encapsulated with the \LaTeX\ 
%\verb+quotation+ environment and is formatted as a single paragraph before the first section heading. 
%(The \verb+quotation+ environment reverts to its usual meaning after the first sectioning command.) 
%Note that numbered references are allowed in the lead paragraph.
%%
%The lead paragraph will only be found in an article being prepared for the journal \textit{Chaos}.
%\end{quotation}

%%%%%%%%%%%%%%%%%%%%%%%%%%% Introduction %%%%%%%%%%%%%%%%%%%%%%%%%%%%%%%
%\section{Introduction} 
{\it Introduction.---}
The \emph{s}-\emph{d} exchange coupling is a basic factor to control spin-dependent transport and magnetization dynamics in ferromagnetic conductors, owing to its ability to transfer angular momentum and energy between conduction spin and magnetization texture.
Among the spintronics effects caused by this quantum mechanical coupling, one of the most extensively studied is spin-transfer torque (STT)\cite{stt1,stt2} that enables an efficient manipulation of the magnetization by an electric current\cite{stt}.
The magnetization dynamics, in turn, can induce an electromotive force (EMF) via the same coupling\cite{Berger,Volovik}, which is known today as spinmotive force (SMF)\cite{smf}.
SMF offers a unique way of detecting dynamical magnetic textures.

Recently, an effect arising from sequential action of STT and SMF in helical magnets has been proposed in Ref.~\cite{Nagaosa2019}, which offers a new principle for an inductor.
When magnetization dynamics is driven by STT due to time-varying electric current, the SMF mechanism leads to an EMF as a counteraction to the applied current, which can, under certain conditions, be interpreted as an inductance.
This inductance of quantum mechanical origin has been coined as emergent inductance\cite{Nagaosa2019}, and experimentally demonstrated in a centrosymmetric helical magnet Gd$_3$Ru$_4$Al$_{12}$\cite{Yokouchi}.
In contrast to a classical inductance of a solenoid coil $ L=\mu n^2 l A $, where $\mu,\,n,\,l,\,A$ are the permeability, turn density, length, and cross-section of the coil, respectively, the emergent inductance $ L_{\rm s} $ is inversely proportional to $ A $. Therefore, the emergent inductor is advantageous for nanoscale device applications as its magnitude increases with decreasing device cross section\cite{Nagaosa2019,Yokouchi}, in contrast to the classical inductance, breaking a hurdle for manufacturing small-size inductors with a large enough effect.

In the meantime, another fundamental and important factor in spintronics, besides the exchange coupling, is spin-orbit couplings (SOCs)\cite{sot}.
The concept of STT has been extended to spin-orbit torque (SOT) by including the effects of SOCs\cite{sot1,sot2}.
While the original STT is already implemented in commercial memory devices\cite{Dieny}, the SOT is expected to play a major role in next-generation technologies\cite{sotMRAM}.
In parallel to these studies on SOT, it has been shown that the SOCs give rise to additional contributions to the SMF\cite{Kim,Tatara,Yamane2013,Yamane2019}.
In particular, helical magnetic textures such as spin spiral are often stabilized by the interplay of symmetric exchange interaction and Dzyaloshinskii-Moriya interaction (DMI), the latter of which stems from the SOC in the electron system.
In those cases, the SOT and SOC-induced SMF are expected to be present.
Furthermore, SOCs are an origin of the electron spin relaxation process, which leads to the so-called $ \beta $-terms in the STT and SMF\cite{Saslow,Duine,Tserkovnyak}.
While these SOC effects sometimes cause drastic change in the spin transport in magnetization textures the previous discussion\cite{Nagaosa2019,Yokouchi} has been limited to the use of adiabatic STT and SMF free from the SOCs.

In this Letter, we scrutinize the role of SOCs in  $ L_{\rm s} $ by taking account of the SOC effects in the STT and SMF processes. 
We identify two key factors that significantly modify $ L_{\rm s} $: 
one is the \emph{intrinsic} origin via the STT and SMF directly augmented by the Rashba effect, and the other is the \emph{extrinsic} origin associated with the spin relaxation and sample disorder. 
The former is intrinsic because it reflects the electronic band structure, and it can be tuned by electrical gating, thus suggesting the possibility of electrical control of $ L_{\rm s} $.
The latter, on the other hand, is extrinsic for the coupling of the spin to the external degrees of freedom, and it is shown to allow $ L_{\rm s} $ to change its sign.
This may partially explain the negative $ L_{\rm s} $ observed in Ref.~\cite{Yokouchi}, while the positive sign had been predicted by Ref.~\cite{Nagaosa2019}.

\begin{figure}[t] %h(here)t(top)b(bottom)p(page)\begin{figure*}
\begin{center}
\includegraphics[width=8cm,clip]{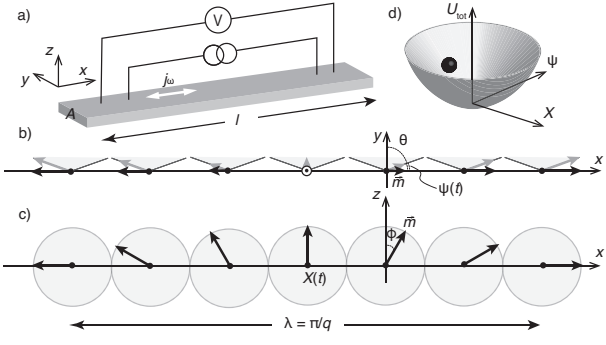}%}
\end{center}
\caption{A set-up of the present study. (a) Schematic view of the device. An ac current $j_\omega$ is applied along the $x$ axis in a spiral magnet thin film with the film normal being along the $z$ axis. The bold arrows show the unit vectors of the magnetization $\vec m$ projected in the $x$--$y$ plane (b), and in the $x$--$z$ plane (c), respectively. The light gray objects indicate the magnetization configuration slightly tilted away from the spiral plane. The Euler angles ($\theta$, $\phi$) and the collective coordinates ($X$, $\Psi$) are also specified. (d) The phase space energy landscape for the ``particle'' (black dot) representing the spiral dynamics. The curvature of the energy originates from the pinning and magnetic anisotropy for $X$ and $\Psi$ respectively.
 }
\label{fig.1} 
\end{figure}

%%%%%%%%%%%%%%%%%%%%%%%%%%%%%%%%
%%%%%%%%%%%%%%%%%%%%%%%%%%%%%%%%
%%%%%%%%%%%%%%%%%%%%%%%%%%%%%%%%
%%%%%%%%%%%%%%%%%%%%%%%%%%%%%%%%
%%%%%%%%%%%%%%%%%%%%%%%%%%%%%%%%
{\it Current-driven dynamics of spiral magnet.---}
We consider a thin-film magnetic strap extending along the $x$ direction, where the magnetization is uniform in the $y$--$z$ plane.
We set the normal direction to the film the $z$ axis as shown in Fig.\ref{fig.1}a.
The broken inversion symmetry at the interface leads to the appearance of DMI.
We assume that the magnetic energy $U$ is given by $ U = \int d^3r \left[ A_\mathrm{s} ( \nabla \vec m )^2 + D \vec m \cdot \left\{ \left( \vec e_z \times \nabla_x \vec e_x \right) \times \vec m \right\} \right] $, 
where $A_\mathrm{s} (>0)$ is the exchange stiffness, $D$ is the DMI constant, $ \vec e_i $ is the unit vector along the $i (= x, y, z) $ axis, and 
$ \vec m = ( \sin\theta \sin\phi , \cos\theta , \sin\theta \cos\phi ) $ is the classical unit vector representing the magnetization direction.
The above form of energy $U$ can stabilize N\'{e}el-type spiral (cycloidal) structures; $\theta=\pi/2$ and $\phi=cqx$, i.e., $
	\vec m ( x )  = \vec e_z \cos \left( c q x \right) + \vec e_x \sin \left( c q x \right) ,
$ as depicted in Fig.\ref{fig.1}b, c with $ c \equiv D/|D| = \pm 1 $ characterizing the chirality of the spiral and the wave number $ q = |D| / 2 A_\mathrm{s}$. 

When an electric current is applied along the $x$ axis, the spiral is driven into motion due to the STT and SOT effects.
We adopt an ansatz that the dynamical spiral structure is obtained by replacing the Euler angles 
$ \phi=cqx \rightarrow cq\left[x - X ( t )\right] $ and $ \theta = \pi / 2 \rightarrow \pi / 2 - \Psi ( t )$\cite{Kishine2,Hals}.
Here we introduce the collective coordinates, $ X $ and $ \Psi $, that describe, respectively, the translational displacement of the spiral along the $x$ axis and tilting away from the easy ($x$--$z$) plane (Fig.\ref{fig.1}b, c), composing a canonical conjugate pair for the dynamical system\cite{Kishine2}. 
Note that $ U = A l K_{\rm ani} \sin^ 2 \Psi $ for the spiral where $ A $ and $ l $ are respectively the cross section area in the $ y $--$ z $ plane and length along the $ x $ axis of the sample (Fig.\ref{fig.1}a). A hard axis anisotropy arises from the DMI as $ K_{\rm ani} = \frac{ D^2 }{ 2 A_{\rm s} }$.

The current-induced magnetization dynamics are in general determined from the Landau-Lifshitz-Gilbert (LLG) equation with STT and SOT included.
Here we focus on the SOT originating from the Rashba SOC and assume that the STT and SOT are derived based on the so-called ferromagnetic Rashba Hamiltonian for a conduction electron given by\cite{ZhangLi,Kim2012}
\begin{eqnarray}
  {\cal H} &=& \frac{{\vec p}^2}{2m_{\rm e}} + J {\vec \sigma} \cdot {\vec m} (x,t)
  + \eta_{\rm R} \left(\sigma_x p_y - \sigma_y p_x\right)  ,
\label{h}
\end{eqnarray}
where $ m_{\rm e} $, ${\vec p}$, and ${\vec \sigma}$ are the electron's mass, canonical momentum, and Pauli matrices of spin operators respectively. The second term is the exchange coupling to the magnetization texture ${\vec m} (x,t)$ with $J(>0)$ the coupling energy and the last term is the Rashba SOC with $ \eta_{\rm R} $ the Rashba parameter. in the unit of velocity. Based on $ U $ and $  {\cal H} $,
a Lagrangian $ {\cal L} $ and a Rayleigh function $ {\cal R} $ for the LLG equation are now given by
\begin{eqnarray}
	{\cal L} &=& \frac{ \mu_0 M_{\rm S} }{ \gamma } \int d^3x \left[\left( 1 - \cos\theta \right) {\cal D}_t \phi 
			+ \gamma\ h_{\rm R} \cos\theta  \right] - U 
  	           , \\
	{\cal R} &=& \frac{ \mu_0 M_{\rm S} }{ \gamma }
                    \int d^3x \left[\frac{ \alpha }{ 2 }\left( {\cal D}_t^\beta \vec m \right)^2
              	   + \beta \gamma h_{\rm R} 
                       \left( \vec m \times \vec e_y \right) \cdot \partial_t \vec m \right], 
\end{eqnarray}
where $ \gamma $ is the gyromagnetic ratio, $ \mu_0 $ is the magnetic constant, $ M_{\rm S} $ is the saturation magnetization, and $ \alpha $ is the Gilbert damping constant.
The STT effects are incorporated in
%\begin{equation}
$	{\cal D}_t = \partial_t + u \nabla_x , $ %\quad
	${\cal D}^\beta_t = \partial_t + \frac{ \beta }{ \alpha } u \nabla_x $ with %\quad
	$u = - \frac{ \hbar \gamma P }{ 2 e \mu_0 M_{\rm S} } j ,$
%\end{equation}
where $ e (> 0 )$ is the elementary electric charge, $P$ is the spin polarization of the electric current, $ \beta $ is a dimensionless constant characterizing the nonadiabatic electron spin dynamics, and $ j $ is the applied electric current density.
The SOT is described by the parameter $ h_{\rm R} $ in the units of a magnetic field, the magnitude of which is evaluated\cite{sot1,Kim} as
%\begin{equation}
$	h_{\rm R} = \frac{ \eta_{\rm R} m_{\rm e} P }{ e \mu_0 M_{\rm S} } j $.
%\end{equation}

We here introduce a pinning potential for translational motion of the spiral, which may originate from disorder, by replacing  $ U \rightarrow U_{\rm tot} = U + U_{\rm pin} $.
For simplicity, we assume for $ U_{\rm pin}$ a quadratic potential around $ X = 0 $, which is justified when the displacement of the spiral is restricted around the pinning position\cite{Thomas};
$U_{\rm pin} =   A l w_{\rm pin} q^3 X^2 /2,$
%\begin{equation}
%	U_{\rm pin} = \frac{ w_{\rm pin} q^3 l }{ \pi } X^2 ,
%\end{equation}
where $ w_{\rm pin} $ is a phenomenological constant. %and $ l $ is the sample dimension in the $x$ direction.
Now the collective coordinates ($X$, $\Psi$) comprise a dynamical system of a particle
trapped in a harmonic potential. Due to $ K_{\rm ani} $, it acquires an effective mass (per unit area and unit length), $ m_{\rm s} = \left( \frac{ \mu_0 M_{\rm S} }{ \gamma } \right)^2 \frac{q^2}{ 2 K_{\rm ani} }$, corresponding to the D\"oring mass of a domain wall\cite{Chikazumi,Saitoh}. Figure \ref{fig.1}d shows the total energy landscape in the phase space spanned by ($X$, $\Psi$) for the present system.
The Euler-Lagrange equations, 
$ \frac{ d }{ d t } \frac{ \delta {\cal L} }{ \delta \dot{ \xi } } - \frac{ \delta {\cal L} }{ \delta \xi }
    = - \frac{ \delta {\cal R} }{ \delta \dot{ \xi } } $ with $\dot{ \xi } \equiv \frac{d \xi }{ d t }$ for $\xi = X$ and $\Psi$,
lead to
\begin{eqnarray}
  - %\frac{ \alpha \lambda }{ \pi } 
  \alpha \partial_t \Psi + c q\cos\Psi \partial_t X
    &=& c q u_{\rm R} \cos\Psi + \frac{ q \nu_{\rm ani} }{ 2 } \sin2\Psi , \label{eom1} \\
  \alpha q \cos\Psi \partial_t X + c \partial_t \Psi
    &=& \beta q u_{\rm R} \cos\Psi - q^2 \nu_{\rm pin} \frac{ X }{  \cos\Psi } . \label{eom2}
\end{eqnarray}
We have introduced %the spatial period of the spiral $\lambda$ and 
the ``velocities,'' $ \nu_{\rm pin} $ and $ \nu_{\rm ani}$, characterizing the pinning strength for $X$ and the anisotropy for $ \Psi $ due to the DMI, respectively, by 
%\begin{equation}
%  \lambda = \frac{ \pi }{ q } , \qquad
 $ \nu_{\rm pin} = \frac{ \gamma w_{\rm pin} }{ \mu_0 M_{\rm S} } $ and % \qquad
  $\nu_{\rm ani} = \frac{ \gamma |D| }{ \mu_0 M_{\rm S} } $.
%\end{equation}
The effects of the electric current are encapsulated in the parameter $ u_{\rm R}  = u - c \gamma h_{\rm R}  / q = \left(  1 + c q_{\rm R}  / q  \right) u, $ %as
%\begin{equation}
%	u_{\rm R} = u - c \gamma h_{\rm R}  / q 
%                         = - \frac{ \gamma \hbar P }{ 2 e \mu_0 M_{\rm S} } \left(  1 + \frac{ c q_{\rm R} }{ q }  \right) j ,
%\end{equation}
with $ q_{\rm R} = 2 m_{\rm e} \eta_{\rm R} / \hbar $.

For $ \Psi \ll 1 $, an approximate solution of Eqs.~(\ref{eom1}) and (\ref{eom2}) in the Fourier form is given by
\begin{equation}
	\left( \begin{array}{c} X_\omega \\ \Psi_\omega \end{array} \right)
     = \left( \begin{array}{cc} i c \omega q^{-1} & \nu_{\rm ani} + i \alpha \omega q^{-1} \\ 
                                           - q \nu_{\rm pin} - i \alpha \omega & i c \omega
        \end{array} \right)
        \left( \begin{array}{c} c \\ \beta \end{array} \right) \frac{ u_{ {\rm R} \omega } }{ \Delta } ,
\end{equation}
where $ \Delta = \left( \nu_{\rm ani} +  i \alpha \omega q^{-1}  \right)
                      \left( q \nu_{\rm pin}  + i \alpha \omega \right)
                      -  \omega^2 q^{-1} $.
When the frequency $ \omega $ is sufficiently low compared with the characteristic to the magnetization dynamics as $ \omega \ll q \nu_{\rm pin}$ and $ \omega \ll  \beta q \nu_{\rm ani} $, the leading terms of the solution read
\begin{align}
X_\omega\simeq \frac{ \beta }{q v_\mathrm{pin}} u_{{\rm R} \omega } , \quad
\Psi_{\omega}\simeq -\frac{c}{v_{\rm ani}} u_{{\rm R} \omega } .
\label{solution}
\end{align}
Note that the above solution is only valid when both $ \nu_{\rm pin} $ and $ \nu_{\rm ani} $ are nonzero. 
Notice also that it is $ X $ and $ \Psi $ themselves, \emph{not their velocities}, that are proportional to the electric current density $ j$.
This implies that a dc current cannot drive persistent magnetization dynamics, where the spiral structure reaches some static state determined by Eq.~(\ref{solution}).
This is essential for the spiral magnet to be interpreted as an ``inductor'' because, as we will see below, the SMF depends on the temporal derivative of the magnetization, which we saw is virtually zero when the electric current is not changing in time.

{\it Electromotive force induced by spiral dynamics.---}
In the following, we discuss the SMF generated by the current-driven spiral dynamics.
First, we summarize the emergent electric field (also known as spin electric field) appearing when the magnetization changes in time. 
Equation~(\ref{h}) leads to the spin electric field $ E^\pm = E_0^\pm + E_{\rm R}^\pm$ along the spiral axis, i.e., the $x$ axis, where the upper (lower) sign corresponds to the electrons with majority (minority) spin\cite{Yamane2019}, and
\begin{eqnarray}
  E_0^\pm  &=&   \pm \frac{\hbar}{2e} 
                                 \left( {\vec m} \times \frac{\partial{\vec m}}{\partial t}
                                         + \beta \frac{\partial{\vec m}}{\partial t}
                                 \right) \cdot \nabla_x {\vec m}  , \label{E0} \\
  E_{\rm R}^\pm &=& \mp \frac{\hbar q_{\rm R} }{2e} 
                                         \left( \frac{\partial {\vec m}}{\partial t} - \beta {\vec m} \times \frac{\partial{\vec m}}{\partial t}\right)_y ,
\label{Eso}
\end{eqnarray}
where $\eta_{\rm R}$ has been assumed to be time-independent\cite{Yamane2013}.
Note that $\beta$ is the same as that appears in the STT reflecting the reciprocal relationship\cite{Duine,Tserkovnyak}.
$E_0^\pm$ is the SOC-free spin electric field, arising when ${\vec m}$ varies in both time and space.
$E_{\rm R}^\pm$, on the other hand, is the SOC-induced part.
In contrast to $E_0^\pm$, the SOC-induced field $E_{\rm R}^\pm$ appears in the spatially uniform ${\vec m}$.

The EMF $ {\cal E} $ due to $ E^\pm $, i.e., SMF, is defined by $ {\cal E} = \int_0^l dx P E^+ $.
The spin polarization $P$ appears because the spin electric field changes its sign when it acts on the majority-spin and minority-spin electrons.
Using the spiral dynamics solution (\ref{solution}) given above, one obtains in the presence of an ac electric current, 
$	{\cal E}_\omega = - \frac{ P \hbar c q l }{ 2 e } \left( 1 + \frac{ c q_{\rm R} }{ q } \right)
                                                   i \omega \left( \Psi_\omega + \beta c q X_\omega \right) 
                                           \simeq  \frac{ P \hbar q l }{ 2 e } \left( 1 + \frac{ c q_{\rm R} }{ q } \right)^2
                                                      \left( \frac{ 1 } { \nu_{\rm ani} }  -   \frac{ \beta^2 }{ \nu_{\rm pin} } \right)
%                                                      \frac{  i \omega u_\omega }{ \nu_{\rm ani} }  ,
                                                      i \omega u_\omega  $.
%\begin{eqnarray}
%	{\cal E}_\omega &=& - \frac{ P \hbar c q l }{ 2 e } \left( 1 + \frac{ c q_{\rm R} }{ q } \right)
%                                                   i \omega \left( \Psi_\omega + \beta c q X_\omega \right) \nonumber \\
%                                           &\simeq&  \frac{ P^2 \hbar q l }{ 2 e } \left( 1 + \frac{ c q_{\rm R} }{ q } \right)^2
%                                                      \left( \frac{ 1 } { \nu_{\rm ani} }  -  \beta^2 \frac{ 1 }{ \nu_{\rm pin} } \right)
%%                                                      \frac{  i \omega u_\omega }{ \nu_{\rm ani} }  ,
%                                                      i \omega u_\omega  .
%\label{smf}
%\end{eqnarray}
%where
%\begin{equation}
% \frac{ \hbar \gamma }{ 2 e \mu_0 M_{\rm S} } j_{\rm ani} = \nu_{\rm ani} .
%\end{equation}
It is clearly seen that the EMF is proportional to the time-derivative of the electric current density.
The spiral magnet, therefore, can behave as an inductor when it is implemented in an electronic circuit.
While the SMF in a chiral helimagnet was first discussed in Ref.\cite{Kishine}, there the SOC effects and $ \beta $-term have not been considered, and the field-induced magnetization dynamics was discussed.

By rewriting $ {\cal E} $ into the form $ {\cal E} = L_{\rm s} dI / dt $ with $ I =  j A $, the emergent inductance is identified as
\begin{equation}
  L_{\rm s} =  \Theta_{\rm int} \Xi_{\rm ext}  L_{\rm s0} ,
%  \frac{ \pi P^2 \hbar l }{ 2 e \lambda A } \left( 1 + \frac{ c q_{\rm R} }{ q } \right)^2
%                                                      \left( \frac{ 1 }{ j_{\rm ani}} -  \frac{ \beta^2 }{ j_{\rm pin} } \right) ,
\label{l}
\end{equation} 
where we defined $ \Theta_{\rm int} = ( 1 + c q_{\rm R} / q ) ^2 $, $ \Xi_{\rm ext} = 1 - \beta ^2 j_{\rm ani} / j_{\rm pin} $, and $L_{\rm s0} = \frac{ \pi P^2 \hbar l }{ 2 e \lambda A j_{\rm ani} } $,
with the spiral half pitch $\lambda =\pi/q$ and two threshold current densities, $ j_{\rm ani} = 2 e | D | / \hbar $, $ j_{\rm pin} =  2 e w_{\rm pin} / \hbar  $, being introduced. %, and $ A $ is the cross section area of the sample; $ I =  j A $.
Equation (\ref{l}) is the key result. 

{\it Tunability of the emergent inductance.---}
Equation~(\ref{l}) reveals an intriguing tunability of the emergent inductance via intrinsic and extrinsic mechanisms of the SOC effects represented by $\Theta_{\rm int}$ and $\Xi_{\rm ext}$, respectively.
The first factor, $ \Theta_{\rm int} $, is directly associated with the Rashba SOC parameter being intrinsic to the electronic band structure of the present inversion broken system\cite{Dresselhaus}.
If one can prepare a spiral of $ q \sim (20 \ {\rm nm})^{-1}$ in spite of $q_{\rm R} \sim (3.5$ \AA $)^{-1}$ evaluated from the prominent value $  \hbar \eta_{\rm R} = 10^{-10} {\rm eV} \cdot {\rm m} $ for the ferromagnetic multilayers\cite{sot2}, $ L_{\rm s} $ will be significantly enhanced by the factor $ \Theta_{\rm int} \sim 3 \times 10^3 $.
Moreover, it is possible to tune the Rashba parameter and thus $ q_{\rm R} $ by electrical gating\cite{sot}.
This tunability can, therefore, invent a field-effect variable inductance.

\begin{figure}[t] %h(here)t(top)b(bottom)p(page)\begin{figure*}
%\rotatebox[origin=lb]{-90}{
\begin{center}
\includegraphics[width=5cm,clip]{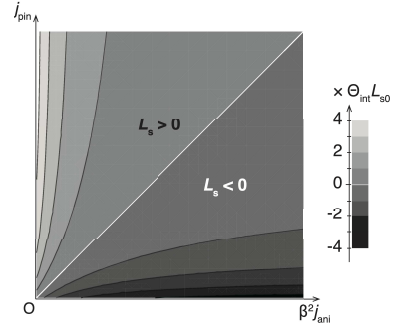}%}
\end{center}
\caption{The contour plot of Eq.~(\ref{l}) in the unit of $ \Theta_{\rm int} L_{\rm s0} $ as a function of $ \beta^2 j_{\rm ani} $ and $ j_{\rm pin} $. The white diagonal line, $\beta = ( j_{\rm pin} / j_{\rm ani} )^{1/2}$, divides positive and negative regions of the emergent inductance $L_{\rm s}$. }
\label{fig.2} 
\end{figure}

The second factor, $ \Xi_{\rm ext} $, arises through the spin relaxation process represented by the $ \beta $-corrections in Eqs.~(\ref{eom2})--(\ref{Eso}) as well as the extrinsic pinning, $ w_{\rm pin} $, being irrespective of the presence of the Rashba SOC.
Interestingly, $L_{\rm s}$ becomes negative for $ ( j_{\rm pin} / j_{\rm ani} )^{1/2} < \beta $ as shown in Fig.\ref{fig.2}.
Although $\beta$ is usually small ($\beta \ll 1$), the condition is not unrealistic as $ j_{\rm pin} / j_{\rm ani} \ll 1 $ can be satisfied by preparing a clean sample with the large-$ K_{\rm ani} $ materials.
While we omit the crystalline anisotropy in this work and $ K_{\rm ani} $ solely comes from the DMI the former would also enhance $ j_{\rm ani} $. 
This may partially explain the recently observed negative emergent inductance\cite{Yokouchi}.

Each effect is factorized in Eq.~(\ref{l}), meaning that the modification does not spoil the attractive devise-size dependence of $ L_{\rm s} $ and that the effects can be tuned independently.

{\it Discussion.---}
Let us compare the essential features obtained above with those of the original form, $ L_{\rm s0} $, derived in Ref.~\cite{Nagaosa2019}, which can be recovered by limiting $q_{\rm R}$, $\beta \to 0$, in Eq.~(\ref{l}).
$ L_{\rm s} $ is inversely proportional to the cross-sectional area, $ A $, as before\cite{Nagaosa2019,Yokouchi}. 
This is because both STT and SMF, and their SOC counterparts as well, are proportional to the electric current density, $ j $, while the classical electromagnetic induction depends on the total current, $ I =  j A $. 
It was claimed in Ref.~\cite{Nagaosa2019} that materials with a shorter spiral pitch, $ 2 \lambda $, (larger $ q $) were favorable to have efficient emergent inductors, providing a guiding principle for experiment\cite{Yokouchi}. This is analogous to the ordinary coil inductance that is proportional to the squared turn number density, $ L \propto n^2 $. 
On the other hand, $ q $ (or $ \lambda $) dependence in Eq.~(\ref{l}) disappears provided that $ q_{\rm R} = \beta = 0 $ since both $ q $ and $ j_{\rm ani} $ originate from the same DMI, canceling out in the present framework, i.e., the result is model-dependent. 
When $ q_{\rm R} $ and $ \beta $ are nonzero $ L_{\rm s} $ exhibits more complex $ q $ dependence. % as discussed below.
The chirality $ c $ just enters in $ \Theta_{\rm int} $ along with the Rashba wavenumber $ q_{\rm R} $. % of Eq.~(\ref{l}). 
If $ q_{\rm R} $ is correlated with the DMI the result does not depend on the chirality and any destructive interference effect due to multi-domain fragmentation is absent as pointed out in Ref.~\cite{Nagaosa2019}.
More specifically, when one follows the chiral derivative approach\cite{Kim2013}, the DMI constant is related to the Rashba SOC as $ D = 2 q_{\rm R} A_{\rm s}$. 
In this case, one arrives at $ q_{\rm R} \equiv c q$, and the universal result $ \Theta_{\rm int} = 4 $. 
To overcome this limit, disentangling $ q_{\rm R} $ with $ q $ is desired.  
Although the relation might be oversimplified
it has been proposed that the DMI constant is proportional to the exchange stiffness\cite{FertLevy,Imamura,Kundu} and that was confirmed experimentally\cite{Nembach}.
The overall sign %in Eq.~(\ref{smf}) 
of $ \cal{E} $
is consistent with the definition of the inductance;
the inductance is positive when the induced EMF, either due to the ordinary electromagnetic induction or the SMF, is opposed to the externally applied voltage.
As clarified above the physical origin of negative $L_{\rm s}$ is ultimately the $ \beta $-correction to the SMF, which has never been observed as a dc response and was omitted in Ref~\cite{Nagaosa2019}.
While the nonadiabaticity and its equivalent contributions arise from many origins\cite{Tatara2007} it is convenient to express it as $ \beta = \frac{\hbar}{ 2 J \tau_{\rm sf}}$, where $ \tau_{\rm sf} $ is the spin flip relaxation time\cite{Yamane2013}.
It can be further correlated to the SOC parameter within the model (\ref{h}) as $1/\tau_{\rm sf} \propto q_{\rm R}^2 \tau $, with $ \tau $ the momentum relaxation time when the D'yakonov-Perel mechanism in inversion symmetry broken systems\cite{Dyakonov} dominates.
The factor $P^2$ in $L_\mathrm{s}$ reflects the fact that the emergent inductor works as a back-reaction to magnetization dynamics induced by applied currents, i.e., it relies on charge-to-spin and spin-to-charge conversions with the efficiency $P$ in each process.
The spin polarization in a slowly varying magnetization texture is well characterized by $ P = \frac{ \sigma_+ - \sigma_- }{ \sigma_+ + \sigma_- } $ with the spin dependent conductivities $ \sigma_\pm $ for majority ($ + $) and minority ($ - $) spin of a uniform system\cite{stt}. The magnitude is roughly estimated by $ P \sim J / \epsilon_{\rm F} $ with $ \epsilon_{\rm F} $ %= \frac{ \hbar^2 k_{\rm F}^2 }{ 2 m_{\rm e} }$ 
the Fermi energy.
It should be noted that the present approach assumes the adiabatic spin transport in slowly varying magnetization and the results are valid for a long wave length regime, $ q \ll  k_{\rm F} $, with $  k_{\rm F} $ the Fermi wave number.
Extending the analysis to an atomistically narrow spiral system, $ q \sim  k_{\rm F} $, is a valuable open problem\cite{Kurebayashi}.

%%%%%%%%%%%%%%%%%%%%%%%%%%%%%%%%
In summary, we have clarified the effects of the Rashba spin-orbit coupling and spin relaxation on the emergent inductor of spiral magnets.
The results are natural extensions of the original work\cite{Nagaosa2019,Yokouchi} while we have unveiled a richer variety of materials dependence that might be tuned by electrical gating, materials design, and careful sample preparation.
Especially, we have pointed out the nonadiabaticity and extrinsic pinning effect are crucial to have a negative emergent inductance.
The knowledge would offer a better understanding of the physics of emergent inductors, propelling further development.

%\begin{acknowledgments}
The authors thank D. Kurebayashi, K. Yamamoto, S. Maekawa, E. Saitoh, S. Fukami, and H. Ohno for fruitful discussions and valuable comments.
This work was supported by JSPS KAKENHI (Grant No. JP16K05424, JP19H05622). 
%\end{acknowledgments}

%\section*{Data availability}
The data that support the findings of this study are available from the corresponding authors upon reasonable request.


%%%%%%%%%%%%%%%%%%%%%%%%%%% References %%%%%%%%%%%%%%%%%%%%%%%%%%%%%%%

{}

\end{document}